\documentclass[prd,groupedaddress, showpacs, twocolumn, amsmath,amssymb,nofootinbib]{revtex4}

\usepackage{graphicx}
\usepackage{dcolumn}
\usepackage{bm}
\usepackage{amsmath,graphics,color}
\usepackage{color,psfrag}
\usepackage{mathrsfs}

\newcommand{\nn}{\nonumber}
\newcommand{\be}{\begin{equation}}
\newcommand{\ee}{\end{equation}}
\newcommand{\bea}{\begin{eqnarray}}
\newcommand{\eea}{\end{eqnarray}}

\newcommand{\vS}{\widetilde{\mbox{\boldmath$\epsilon$}}}

\begin{document}

\def\la{\langle}
\def\ra{\rangle}
\def\a{\alpha}
\def\b{\beta}
\def\g{\gamma}
\def\d{\delta}
\def\D{\Delta}
\def\e{\epsilon}

\def\phi{\varphi}
\def\k{\kappa}
\def\m{\mu}
\def\n{\nu}
\def\o{\omega}
\def\O{\Omega}
\def\p{\pi}
\def\r{\rho}
\def\s{\sigma}
\def\S{\Sigma}
\def\c{\chi}
\def\x{\xi}
\def\t{\tau}
\def\del{\partial}
\def\nab{\nabla}

\title{Gedanken experiments on nearly extremal black holes and the Third Law}

\author{Goffredo Chirco${}^1$}
\email{chirco@sissa.it}
\author{Stefano Liberati${}^1$}
\email{liberati@sissa.it}
\author{Thomas P. Sotiriou${}^2$}
\email{T.Sotiriou@damtp.cam.ac.uk}
\affiliation{${}^1$SISSA-International School for Advanced Studies,  Via Bonomea 265, 34136 Trieste, Italy and INFN sezione di Trieste}
\affiliation{${}^2$Department of Applied Mathematics and Theoretical Physics, Center for Mathematical Sciences, University of Cambridge, Wilberforce Road, Cambridge, CB3 0WA, UK}

\begin{abstract}
A gedanken experiment in which a black hole is pushed to spin at its maximal rate by tossing into it a test body is considered. After demonstrating that this is kinematically possible for a test body made of reasonable matter, we focus on its implications for black hole thermodynamics and the apparent violation of the third law (unattainability of the extremal black hole). We argue that this is not an actual violation, due to subtleties in the absorption process of the test body by the black hole, which are not captured by the purely kinematic considerations. 
\end{abstract}
\pacs{04.70.Dy, 04.70.Bw}
\maketitle

\section{Introduction}
The stationary, vacuum black hole solution of Einstein's equation, called the Kerr spacetime, is unique. It is characterized {by} two parameters, $M$ and $J$. When $M^2\geq J$ (we use units where $G=c=1$), the Kerr spacetime describes a black hole of mass $M$ and angular momentum $J$. When $J=M^2$ the black hole is said to be extremal, as it is spinning at its maximal rate. For $J>M^2$ the Kerr metric has no horizon and has a naked singularity.

It has recently been shown in Ref.~\cite{jaso}, following earlier similar studies and results \cite{Wald,Hubeny:1998ga,Hod:2002pm,deFelice:2001wj}, that one might be able to make a nearly extremal black hole ``jump" over extremality by tossing into it a test body composed of reasonable matter which satisfies energy conditions. This might constitute a violation of the cosmic censorship conjecture, according to which naked singularities cannot be formed via any process that involves physically reasonable matter \cite{Penrose:1969pc}. As radiative and self force effects are neglected in such arguments, it is conceivable, and perhaps expected considering the existing evidence for Cosmic Censorship, that overspinning will be precluded by such effects \cite{jaso}. This makes their study more pertinent.

On the other hand, one can ask what are the implication of the aforementioned result on black hole thermodynamics. Hawking's has provided a proof for the second law of black hole mechanics \cite{Hawking:1971tu}, that the black hole horizon area cannot decrease, and Israel a proof for the third law \cite{is}, that the surface gravity cannot be reduced to zero at a finite time. Turning a black hole into a naked singularity appears to invalidate both laws. However, since in such a process the final object is not a black hole, it is not possible to analyze the thermodynamical implications any further.

Nonetheless, it is intuitive that if one can overspin a nearly extremal black hole, then one should also be able to make it exactly extremal via the same process. In this case, the final object would still be a black hole and the process would appear to violate the third law as it is expressed in Israel's formulation:

\emph{``A nonextremal black hole cannot become extremal at a finite advanced time in any continuous process in which the stress energy tensor of accreted matter stays bounded and satisfies the weak energy condition in a neighborhood of the outer apparent horizon"}~\cite{is}.

We first show, following Ref.~\cite{jaso}, that, at least at the level where radiative and self force effects are neglected, one can indeed drive a nearly extremal black hole to extremality by tossing in a test body. We then proceed to discuss whether this can indeed be considered a violation of the third law of black hole thermodynamics.

\section{Spinning a Kerr black hole to extremality with a test body}\label{spinningup}

We start by considering an uncharged stationary Kerr black hole solution and we attempt to make it exactly extremal by tossing into it a body which carries angular momentum.  Given the \emph{uniqueness theorem} for the Kerr spacetime as a rotating black hole solution and the scale independence of the problem at hand, the angular momentum to mass ratio is an effective control parameter for the family of uncharged stationary Kerr black hole solutions.

The body tossed into the black hole is treated as a test body with energy $\d E$ and the angular momentum $\d J$, such that
\be
\d E\,,\,\d J\ll M. \label{i}
\ee
The term test body is used here to mean a body whose energy and angular momentum satisfy Eq.~(\ref{i}) so that radiative and self force effects can be neglected and the body can be considered as moving on a Kerr background.

Following Ref.~\cite{jaso}, one can imagine a process by which a series of test body absorptions pushes the Kerr black hole to become initially nearly extremal and then exactly extremal (instead of over extremal, which was the case considered in Ref.~\cite{jaso}). Clearly, this final step to exact extremality will be the crucial one. The test body quantities $\d E$ and $\d J$ would have to satisfy
\be
J+\d J= (M+\d E)^2. \label{j}
\ee
For a generic $\d E$ Eq.~(\ref{j}) can be considered as fixing the required angular momentum carried by the test body, namely
\be
\d J=(M^2-J)+2M\d E+\d E^2. \label{jex}
\ee
Note that from Eq.~(\ref{jex}) one can infer that
\be
\label{size}
\d J/\d E^2>2M/\d E\gg 1
\ee
which implies that the test body cannot be itself a black hole. This in turn imposes the restriction that the test body has a finite size which should be bigger than its Schwarzschild radius, even if its angular momentum is orbital and not due to spin.

A second condition on the test body energy and momentum will come by requiring that the test body indeed crosses the horizon of the black hole.  The following relation holds 
\be
\d E-\O_H \d J=\int T_{ab} {\c}^a d\S^b, \label{cons}
\ee
and can be considered as giving the flux of energy and angular momentum into the black hole, where $T_{ab}$ is the stress energy tensor of the body,  $\O_H=a/2Mr_+$ is the horizon angular velocity, $a=J/M$ is the specific angular momentum and $r_+=M+(M^2-a^2)^{1/2}$ defines the horizon radius in Boyer-Lindquist coordinates. The quantity $\c^a=\del_t^a+\O_H\del_{\phi}^a$ defines the horizon-generating Killing vector, while $d \S^b$ is the horizon surface element.

Assuming that the matter of which the test body is composed satisfies the null energy condition (NEC) implies that the flux of energy and angular momentum is positive and
Eq.~(\ref{cons}) yields
\be
\d J \le \d E/\O_H\equiv \d J_{max} \label{bond}.
\ee
This equation  could have been derived also by analyzing the trajectories using the geodesic or the Papapetrou equations, depending on whether the test body's angular momentum is orbital or due to spin.
  
Given that $\d J$ satisfies Eq.~(\ref{jex}), Eq.~(\ref{bond}) turns into a second order inequality for $\d E$,
\be
\O_H\d E^2+(2M\O_H-1)\d E+ \O_H(M^2-J) \le 0, \label{eq}
\ee
whose determinant $\D=4\O_H(\O_HJ-M)+1$ depends only on the black hole quantities and is independent of the test body parameters. One can verify numerically that
\bea
\D=0\,\,\, &\iff& \,\,\, J=M^2\\
\D >0\,\,\, &\forall&\,\, \,J<M^2. \label{cond2}
\eea
Clearly, the case where $J=M^2$ is not interesting for this discussion, as it is the case where the black hole is already extremal initially.
On the other hand, Eq.~(\ref{cond2}) implies that as long as the black hole starts in a nonextremal configuration, one can always find values for $\d E$ and $\d J$, for which both  Eqs.~(\ref{jex}) and~(\ref{bond}) are satisfied. 

Before concluding that a black hole can be driven to extremality via the gedanken experiment discussed we also need to consider size and structural constraints on the test particles. Such requirements can be imposed by requiring that the particle is composed by reasonable matter, which does not violate energy conditions. Such a discussion requires that we distinguish between two different cases which we were able to treat on equal footing so far: the case where the angular momentum is due to spin and the one were it is purely orbital. 

An analysis of both cases has been already performed in Ref.~\cite{jaso} in an attempt to overspin a Kerr black hole. The only difference from our scenario is that Eq.~(\ref{jex}) is now an equality instead of {an} inequality. However, this can hardly affect the end result, as one can easily verify by attempting to trace the various steps of the argument.  It is actually intuitive that if a body can have enough angular momentum for a given energy to be able to overspin a black hole without violating energy conditions then there is no reason why a body cannot have the right angular momentum to make it exactly extremal, as the latter is actually less. So, we avoid repeating the analysis here and we state the outcome, referring the reader to Ref.~\cite{jaso} for further details. 

For a particle with spin,  the size and structural constraints require that it has to be deeply bound and somewhat oblate, otherwise it cannot carry the right amount of angular momentum and energy. On the other hand, if the angular momentum is orbital then the particle can very well be on an unbound trajectory, i.e. fall into the black hole from infinity.

In conclusion, we have established that, in the test body approximation, it is possible to drive a Kerr black hole to extremality by tossing in a test particle with the right energy and angular momentum. This in turn seems to imply that a violation of the third law of the black hole thermodynamics is possible via such a process.

\section{Test body absorption and black hole thermodynamics}

As mentioned earlier, it is conceivable that radiative and back reaction effects which have been neglected here could change the kinematics of the process considered above, possibly preventing the test body from falling into the black hole in the first place. Moreover, in the presence of these effects the spacetime would not be stationary anymore, so any reference to black hole thermodynamics could be debatable.

However, we do not think that such objections should be used to dismiss concerns about the possible violation of the third law of black hole thermodynamics. First of all, it is worth establishing if there is an actual violation even within the test body approximation in the first place, before resorting to more complicate physics. Second, one should be able to give a clear answer to whether such a gedanken experiment can fall within the purview of black hole thermodynamics, and indeed this could be ultimately related to the fact that radiative and back reaction effects become important.

So, sticking to the test body approximation, let us assume, as a starting point, that the black hole's stationary properties are not compromised and that its temperature is always well defined. This then implies that the black hole zeroth law holds, even at the extremal limit, where the horizon is degenerate and the temperature is zero.\footnote{However, see Ref.~\cite{Preskill:1991tb, bel} for subtleties and limitation regarding the statistical description of extremal black holes.}

The first and second laws appear to be respected as well. Indeed, Eq.~(\ref{cons}) is nothing but an energy conservation condition, while the NEC actually guaranties that the black hole area change associated to the energy income is always positive. In the approach to extremality the area does not decreases. Indeed, in the $a\to M$ limit, the black hole horizon would shrink if the black hole mass was kept constant along the process. However, the test body absorption increases the black hole mass and the black hole radius actually increases.\footnote{The fact that our approach never violates the area law is not in contrast with a possible violation of third law. The Israel theorem should be associated to the third law of thermodynamics in its weak formulation, which states that it is impossible to reach absolute zero temperature in a physical process. The strong version of the third law of thermodynamics, which states that as the temperature approaches zero, the entropy also approaches zero, does not have an analogue for black holes.}

On the other hand, we have seen that within the same framework (not necessarily correct as we will see shortly), the third law appears to be violated, as an extremal black hole appears to be created by tossing physically reasonable matter in a nearly extremal black hole.
Therefore, the implications of the gedanken experiment discussed here on the third law deserve closer investigation.

We can start by noticing that the test body approach does not necessarily satisfy all of the assumptions contained in Israel's theorem formulation.\footnote{Note that we always refer to the formulation of the third law as stated in Israel's theorem. From a purely thermodynamical point of view, the ``unattainability'' principle of the third law does not  necessarily require equilibrium or adiabatic linkages \cite{Lan,bel}. The Israel's formulation is weaker in this sense. However, since the black hole systems are stationary equilibrium solutions of Einstein's equation, we always deal with equilibrium thermodynamics.}  First of all, in our previous kinematic considerations we assumed that the NEC holds, which is a weaker assumption than the weak energy condition (WEC) required in the third law.
While WEC$\Rightarrow$NEC, NEC$\nRightarrow$WEC. In this sense, by assuming the NEC, we can not  be sure the test body approach is not somehow violating the WEC.

However, given that the test body has positive energy density when released from infinity and that structural constraints are satisfied, there is no reason to believe that the WEC will be violated. Therefore, this issue does not seem to play the crucial role here.

A more fundamental point is related to the fact that the notion of finite advanced time used in the Israel theorem is not clearly addressed in the test body approach. Actually, in the test body approximation, the only time coming into play is the one the particle takes in order to reach the horizon surface along its trajectory. One can indeed find a broad class of trajectories by which the test body with the right energy and angular momentum reaches the horizon in a finite amount of time. However, the time scale considered in the Israel third law formulation is the net time required by the black hole to absorb the particle energy and angular momentum. In the test body approximation this process is ideally instantaneous. 
Note that the test body has a finite size, see discussion after Eq.~$(\ref{size})$, so it will actually take a nonzero amount of proper time to cross the horizon, something that is not captured in the test body approach. 

In order to address this problem, one could try to switch to a thermodynamical description and consider the test body absorption process as a slight perturbation of the black hole equilibrium state. 

Within a classical thermal framework, one possible way to relate the energy transferred to a generic system with some rate of entropy transfer is given by the Bekenstein bound relation~\cite{be,hod},
\be
\D S/\t \le\pi \D E/\hbar \label{bb1},
\ee 
where $\D E$ measures the total energy transferred to the system and $\t$ indicates the  time necessary for the total energy absorption. Moreover, the assumption of local equilibrium condition, 
\be
\D E=T \D S \label{cla},
\ee 
allows one to rewrite the Bekenstein bound as~\cite{hod},
\be
\t \ge \hbar/\pi T \label{bb},
\ee
thereby providing a lower bound on the thermal system relaxation time. 

Of course, the Bekenstein bound has not been rigorously proven. So, its use, or the use of the corresponding relaxation time, should be viewed skeptically. Furthermore, the four laws of black hole dynamics do not rely on quantum mechanics \cite{BCH} (their interpretation as thermodynamical laws, however, does require the introduction of the Hawking temperature and, hence, implicitly relies on quantum physics). Therefore, it might seem odd that one needs to refer to a quantum based bound, such as  Bekenstein's, in order to investigate possible violations of the third law. Indeed, there is no such need. Given the thermal characterization of our absorbing black hole, we simply choose to use the Bekenstein bound argument for orientation purposes, {\em i.e.}~as a tool for gaining some intuition about where the subtlety in the whole process might be. At the end, our claims will not rest on its validity.

Assuming now that black hole thermodynamics holds up to the extremal limit, one could apply the previous argument to the black hole as a thermodynamical system. In this case $\D E$ can be identified with the heat associated to the test body absorption, 
\be
\d Q=\int T_{ab} {\c}^a d\S^b \label{q},
\ee
 and
the temperature $T$ will be the Kerr black hole Hawking temperature $T=\hbar \k/2 \pi$, with surface gravity $\k$ given by
\be
\k=\frac{(M^2-a^2)^{1/2}}{2M^2+2M(M^2-a^2)^{1/2}} \label{kk}.
\ee
The time scale $\tau$ can then be interpreted as the black hole relaxation time (in the black hole reference frame), that is the time interval required by the perturbed black hole solution to recover its thermal equilibrium state, thereby characterizing the transition between two subsequent equilibria.
In the gedanken experiment considered here, Eq.~(\ref{bb}) can be associated to some measure of the time elapsing, in the black hole reference frame, between the arrival of the test body front and the instant the test body back passes the horizon and allows us to get the estimation we need on the time required by the test body satisfying the size and kinematical constrains to complete the transfer process described in the Israel theorem.

Now, a direct application of Eq.~(\ref{bb}) to the previous test body absorption setting would give
\be
\t \ge \hbar/\pi T^{NE} \label{obb},
\ee  
where $T^{NE}$ is the initial temperature of the nearly extremal black hole. Since $T^{NE}$ is always a finite quantity, the relation above indicates that the transfer process can be achieved in a finite time. Therefore, there would be a third law violation. 

On the other hand, given Eqs.~(\ref{bb}) and~(\ref{kk}), one easily realizes that the closer to the extremal limit the black hole solution starts the greater the time to recover its equilibrium will become. This immediately raises doubts about whether one should be using initial nearly extremal black hole equilibrium solution quantities $(M,\O,T)_{NE}$ in order to calculate the relaxation time. In fact, this would implicitly assume that the temperature stays  constant throughout the process. 

However, as the test body transfer of energy and momentum to the black hole takes place, the system temperature moves from $T^{NE}$ to $T=0$, and the temperature change with respect to the angular momentum to mass ratio,
\be
\frac{d T}{da}=-\frac{\hbar\,a}{4\pi\,\sqrt{M^2-a^2}\left(M+\sqrt{M^2-a^2}\right)^2}, \label{der}
\ee
actually diverges as $(M-a)^{-1/2}$ in the $T\to0\,\,\,(a\to M)$ limit. Even though for large enough $T$ the temperature change can be small enough so that the temperature can be considered constant throughout the process of absorption, this is definitely not the case when $T\to 0$. Note that this observation does not hinge on the use of the Bekenstein bound.

To get a deeper insight into the problem, let us entertain the idea that we can describe the test body absorption and the  black hole transition from nearly extremal to extremal as a continuous succession of equilibrium states. As a minimal trial, one can assume that along the transition 
\be 
\d Q=\int _{\D H}T \d S \label{intcla},
\ee
where now both temperature and entropy changes are taken into account. Given the nonzero test body size, the absorption can be thought to effectively involve a section $\D H$ of the Kerr outer horizon null hypersurface, with extrema given by the values of the Killing parameter $v$ at which the front end of the test body crosses the horizon. In this sense, the integral above is ideally taken over the whole test body absorption process. More quantitatively, relation (\ref{intcla}) can be rewritten as
\be
\int_{\D H} T_{ab} \c^a d\S^b=\,\int_{\D H} T\,s^a\,d\S_a. \label{vintcla}
\ee
the temperature being a function of the Killing parameter, $T=\k_v/2\pi$, while $s^a$ defining the matter entropy flux \cite{fla,bwa}. In order to describe the process as a continuous transition between a series of equilibrium states, one actually need the integral relation in~(\ref{intcla}) to be satisfied pointwise along the horizon, that is one needs to further assume 
\be
T_{ab} \c^a \c^b=\frac{\k_v}{2\pi}\,s^a\c_a, \label{locla}
\ee
where we used $d\S^a=\vS\, dv \,\c^a$ in relation~(\ref{vintcla}). In this case, $T_{ab} \c^a \c^b$ indicates the transferred energy density, while $s^a\c_a$ measures the local entropy density. 

Equation~(\ref{locla}) allows to discretize the single test body absorption in a series of smaller processes, by splitting the integral over a sum of subintervals $\d H$, provided that over each interval the \emph{adiabatic condition}
\be
\left|\frac{\dot{\k}}{\k^2}\right|\ll1, \label{ad}
\ee
holds, that is asking for the temperature to stay almost constant at each step. The dot here indicates the derivative with respect to the proper time $v$.
Within these assumptions, one could then define the discrete quantities
\begin{eqnarray} 
\d Q_i &\equiv& \int_{S^2} \int_{v_{i-1}}^{v_i} \vS\, dv\, T_{ab} \c^a \c^b\\ \nn
\tilde{T}_i\,dS_i &\equiv& \tilde{T}_i \int_{S^2} \int_{v_{i-1}}^{v_i} \vS\, dv\,\,s^a\,\c_a, \label{tdsi}
\end{eqnarray}  
with $\tilde{T}_i\equiv(T_{i-1}+T_{i})/2$, and rewrite relation~(\ref{intcla}) as
\be
\sum_{i} \d Q_i=\sum_{i} \tilde{T}_i\, dS_i. \label{parti}
\ee

{If one want to return to a discussion in terms of the Bekenstein bound, then for each i-step, the latter} would still regulate the partial entropy $dS_i$ rate of transfer, that is
\be
d S_i/\t_i \le\pi \d Q_i/\hbar \label{ibb},
\ee 
and, given relation~(\ref{locla}), one would recover
\be
\t_i \ge (\hbar/\pi)/ \tilde{T}_i\label{iobb},
\ee 
where now $\t_i$ measures the time lapse $\D v=v_i-v_{i-1}$ along the horizon. Given Eq.~(\ref{iobb}), one could eventually define a total finite time for the absorption process as the sum
\be
\t=\sum_{i}\t_i =(\hbar/\pi)\sum_{i}\frac{1}{ \tilde{T}_i}\label{sum}.
\ee
This agrees with the intuitive discussion presented earlier. It is now clear that how large $\tau$ is depends on how large can the last step of this process be and still be considered a single step, in the sense of the temperature being almost constant. 

However, the whole discussion about the Bekenstein bound and the related relaxation time can be easily omitted now. One can instead focus on the adiabatic condition, {which can be derived directly from the Raychaudhuri equation (see Ref. \cite{tedrenaud}).} We need to translate Eq.~(\ref{ad}) in terms of the quantities appearing in Eq.~(\ref{der}), that is of quantities which we can actually measure. The adiabatic condition can take the form

\be
\left|\frac{1}{\k^2}\,\frac{d\k}{da}\right| \ll\,\left|\frac{dv}{da}\right|. \label{acc}
\ee
Using Eq.~(\ref{kk}) this inequality becomes
\be
\frac{2M^2a}{(M^2-a^2)^{3/2}} \ll \left|\frac{dv}{da}\right|. \label{acc1}
\ee

In the extremal limit, $a\to M$, the left hand side diverges as $(M-a)^{-3/2}$. This implies that the process could be adiabatic only if the right hand side diverges as well. However, if this would be the case, then it would take an infinite time to reach $a=M$ as $dv/da$ is the rate of change of the affine parameter with respect to $a$. 
What this means is that the process could only be adiabatic if it is infinitely slow, or that if the absorption is to take place in a finite time it cannot be approximated by a succession of equilibria. In this sense, we conclude that the third law \cite{is} cannot be violated by the process described in Sec. \ref{spinningup}. This result is in agreement with earlier studies on the topic \cite{bel, dad, Ro}.

\section{conclusions}

It has been shown that, based on purely kinematic considerations and within the test body approximation, it appears to be possible to create an extremal black hole by tossing a body with suitable energy and angular momentum into a nearly extremal one. The test body can satisfy size and structure constraints imposed by the requirement that it be composed by physically reasonable matter.

This might initially appear as a violation of the third law of black hole thermodynamics, expressed as the unattainability of an extremal black hole in Israel's theorem. We have shown, however, that the absorption of such a test body by a nearly extremal black hole cannot be described as a succession of equilibrium states unless the absorption is assumed to last infinitely long. 

In this sense, even though the test body can reach the horizon at a finite proper time, a extremal black hole cannot be created in a finite time in a continuous process. Therefore, the third law of black hole thermodynamics is not actually violated.

\begin{acknowledgments} The authors would like to thank T. Jacobson and F. Belgiorno for a critical reading of this manuscript and useful comments towards its improvement.
T.P.S. was supported by STFC.
\end{acknowledgments}

\end{document}